\documentclass[12pt]{article}
\usepackage{amsmath,amssymb,amsbsy,latexsym}

\def\parder#1{ \frac{\partial}{\partial #1 }}
\def\odder#1{ \frac{d}{d #1 }}

\global\arraycolsep=1pt
\def\fnote#1#2{\begingroup\def\thefootnote{#1}\footnote{#2}\addtocounter
{footnote}{-1}\endgroup}

\newcommand{\beq}{\begin{equation}}
\newcommand{\eeq}{\end{equation}}
\newcommand{\beqa}{\begin{eqnarray}}
\newcommand{\eeqa}{\end{eqnarray}}

\newcommand{\CS}{{\mathcal S}}
\newcommand{\CL}{{\mathcal L}}
\newcommand{\CN}{{\mathcal N}}
\newcommand{\CR}{{\mathcal R}}

\begin{document}
\begin{flushright}
OCU-PHYS 230 \\
OIQP-05-07\\
hep-th/0505259\\
\end{flushright}
\vspace{8mm}

\begin{center}
{\bf\Large
Scalar Laplacian on Sasaki-Einstein Manifolds $Y^{p,q}$
}

\vspace{10mm}
Hironobu Kihara\fnote{$*$}{
\texttt{kihara@sci.osaka-cu.ac.jp}
}\\
\textit{\small
Osaka City University Advanced Mathematical Institute (OCAMI)\\
3-3-138 Sugimoto,
Sumiyoshi,
Osaka 558-8585, Japan
}\vspace{4mm}

~
Makoto Sakaguchi\fnote{$\dag$}{
\texttt{makoto\_sakaguchi@pref.okayama.jp}
}\\
\textit{\small
Okayama Institute for Quantum Physics\\
1-9-1 Kyoyama, Okayama 700-0015, Japan
}
\vspace{4mm}

Yukinori Yasui\fnote{$\ddag$}{
\texttt{yasui@sci.osaka-cu.ac.jp}
}\\
\textit{\small
Department of Mathematics and Physics,
Graduate School of Science,
Osaka City University\\
3-3-138 Sugimoto,
Sumiyoshi,
Osaka 558-8585, Japan
}

\end{center}
\vspace{8mm}

\begin{abstract}
We study the spectrum of the scalar Laplacian
on the five-dimensional toric Sasaki-Einstein
manifolds $Y^{p,q}$.
The eigenvalue equation reduces to Heun's equation,
which is a Fuchsian equation with four regular singularities.
We show that the ground states,
which are given by constant solutions of Heun's equation,
are identified with BPS states corresponding to
the chiral primary operators in the dual quiver gauge theories.
The excited states
correspond to non-trivial solutions of Heun's equation.
It is shown that these  reduce to polynomial solutions
in the near BPS limit.

\end{abstract}
\newpage

The AdS/CFT correspondence \cite{AdS/CFT}
has attracted much interest
as a realization of the string theory/gauge theory correspondence.
It predicts that
string theory in AdS$_5\times X_5$
with $X_5$ be Sasaki-Einstein
is dual to $\CN=1$ 4-dimensional superconformal field theory.
Recently in \cite{Ypq in 5-dim,toric SE},
5-dimensional inhomogeneous toric Sasaki-Einstein manifolds $Y^{p,q}$
are explicitly constructed,
besides the homogeneous manifolds,
S$^5$ and $T^{1,1}$.
The associated Calabi-Yau cones $C(Y^{p,q})$ are toric
owing to the presence of the $T^2$-action
on the 4-dimensional K\"ahler-Einstein bases.
Thanks to this property,
the authors of \cite{BBC}\cite{quiver}
clarified
the $\mathcal{N}=1$ 4-dimensional dual superconformal field theories
of IR fixed points of toric quiver gauge theories.
(Further developments in this subject include \cite{further}\cite{BHOP}.)
On the other hand, in the gravity side,
semiclassical strings moving on the AdS$_5\times Y^{p,q}$ geometry
are shown to be useful to establish the AdS/CFT correspondence 
in \cite{semiclassical string}.

In this letter, 
we study the spectrum of the scalar Laplacian on the
5-dimensional Sasaki-Einstein manifolds $Y^{p,q}$.
The eigenvalue equation of the scalar Laplacian
is shown to reduce to Heun's equation\fnote{$\sharp$}{
In \cite{Gibbons:2004em}, Heun's equation corresponding to
the scalar Laplacian on the inhomogeneous
manifolds constructed in \cite{Hashimoto:2004kc}
is examined. 
}
after the separation of variables.
Heun's equation is the general second-order linear Fuchsian equation with
\textit{four} singularities.
It is known that the methods to investigate hypergeometric functions
with three singularities do not work for Heun's equation.
Though there exist power-series solutions,
the coefficients are governed by three-term recursive relations,
and thus it is generally impossible 
to write down these series explicitly.
We clarify some eigenstates of the scalar Laplacian,
i.e. solutions of Heun's equation,
which include BPS states
dual to chiral primary operators of the superconformal gauge theory.

\medskip

The metric tensor of $Y^{p,q}$ 
parameterized by two positive integers $p,q$ $(p> q)$ is
written as~\cite{Ypq in 5-dim}
\begin{eqnarray}
ds^2
&=&\frac{1-y}{6} \left( d \theta^2 + \sin^2 \theta d \phi^2
		      \right) + \frac{1}{w(y)q(y)} dy^2 +
\frac{q(y)}{9}( d \psi - \cos \theta d \phi  )^2 ~~\nonumber\\ 
&& + w(y) \left[ d \alpha + f(y)(d \psi -
	   \cos \theta d \phi)   \right]^2~~,
\label{eqn:ypqmetric}
\end{eqnarray}
with 
\begin{eqnarray}
w(y) &=& \frac{2(b - y^2)}{1-y} ~,~~~
q(y) =\frac{b-3y^2 +2y^3}{b-y^2}~,~~~
f(y) = \frac{b - 2y + y^2}{6(b - y^2)}~,\nonumber\\
b &=&  \frac{1}{2} - \frac{p^2 -3 q^2}{4p^3} \sqrt{4p^2 - 3 q^2}~.
\end{eqnarray}
The coordinates $\{y,\theta,\phi,\psi,\alpha\}$ have the following ranges:
\begin{eqnarray}
y_1 \leq y \leq y_2~,~~~ 0 \leq \theta \leq \pi ~,~~~
0 \leq \phi \leq 2 \pi~,~~~  0 \leq \psi \leq 2 \pi~,~~
0 \leq \alpha \leq 2 \pi l ~. 
\end{eqnarray}
The boundaries $y=y_1, y_2$ are given by the two smallest roots of the
cubic $b-3y^2 +2y^3$, 
\begin{eqnarray}
y_{1,2} = \frac{1}{4p} \left( 2 p \mp 3q - \sqrt{ 4 p^2 - 3 q^2 }
			 \right) ~~,
\end{eqnarray}
respectively,
while the remaining root takes the value 
\begin{eqnarray}
y_3 &=& \frac{3}{2} -(y_1 + y_2)
= \frac{1}{2} + \frac{\sqrt{4p^2 -3q^2}}{2p} ~~.
\end{eqnarray}
The period of $\alpha$ is $2 \pi l$ with
\begin{eqnarray}
l = \frac{q}{3 q^2  - 2 p^2 + p \sqrt{ 4 p^2 - 3 q^2 }}~~.
\end{eqnarray}

For the metric (\ref{eqn:ypqmetric})
we have the scalar Laplacian\fnote{$\natural$}{
The spectrum of the scalar Laplacian on $T^{p,q}$
is examined in
 \cite{Gubser:1998vd}.
For that of the scalar Laplacian on the Calabi-Yau cone $C(Y^{p,q})$,
some properties are studied in \cite{BHOP}.
},
\begin{eqnarray}
\square  &=& 
\frac{1}{1 - y} \parder{y} (1-y) w(y)q(y) \parder{y} \nonumber\\&&
+ \left( \frac{3}{2} \hat{Q}_R \right)^2 
+ \frac{1}{w(y)q(y)} \left(\parder{\alpha } + 3y \hat{Q}_R \right)^2  
+ \frac{6}{1-y}\left[ \hat{K} - \left( \parder{\psi}  \right)^2 \right]~.
\end{eqnarray}
The operator $\hat{Q}_R$ represents the Reeb Killing vector field,
\begin{eqnarray}
\hat{Q}_R &=& 2 \parder{\psi} - \frac{1}{3} \parder{\alpha}~~,
\end{eqnarray}
which corresponds to the ${\mathcal R}$-symmetry of the dual gauge theory
\cite{quiver}.
The operator $\hat{K}$ is the second Casimir of $SU(2)$,
which is a part of the isometry $SU(2)\times U(1)^2$,
\begin{eqnarray}
\hat{K} = \frac{1}{\sin \theta} \parder{\theta} \sin \theta
 \parder{\theta} + \frac{1}{\sin^2 \theta} \left(\parder{\phi} + \cos
 \theta \parder{\psi}  \right)^2 + \left( \parder{\psi} \right)^2 ~.
\end{eqnarray}
Owing to the isometry,
the eigenfunction takes the form
\begin{eqnarray}
\Phi(y,\theta,\phi,\psi,\alpha) &=& \exp\left[ i \left( N_{\phi} \phi +
 N_{\psi} \psi + \frac{N_{\alpha}}{l} \alpha  \right) \right]
  R(y) \Theta(\theta)~
\end{eqnarray}
with $N_{\phi} , N_{\psi} , N_{\alpha} \in {\mathbb Z} $.
Then the equation,
 $\square \Phi = - E \Phi$,
reduces to
\begin{eqnarray}
\hat{K} \Theta(\theta) = - J(J+1) \Theta(\theta) ~~,
\label{eqn:odefortheta}
\end{eqnarray}
and
\begin{eqnarray}
&&\frac{1}{1-y}  \frac{d}{dy}\left[ (1-y) w(y)q(y) \frac{d}{dy}
				 R(y)\right]
 - \left[\left(\frac{3}{2} Q_R \right)^2
 +
\right. 
\nonumber \\&&
  \frac{1}{w(y)q(y)} \left( \frac{N_{\alpha}}{l} + 3 y Q_R  \right)^2  
\left. 
 +
\frac{6}{1-y}\left( J(J+1) - N_{\psi}^2  \right) - E\right] R(y) = 0
~.
\label{eqn:odefory}
\end{eqnarray}
The quantum numbers $J$ and $Q_R = 2 N_{\psi} - \frac{1}{3l}N_{\alpha}$
correspond to $SU(2)$-spin and ${\mathcal R}$-charge, respectively.
The regular solutions of
the first equation (\ref{eqn:odefortheta}) are given by Jacobi polynomials.
After some calculations 
we find that the second equation (\ref{eqn:odefory}) is
of Fuchsian-type with four regular singularities at
 $y=y_1, y_2, y_3$ and $\infty$, i.e. Heun's equation;
\begin{eqnarray}
\frac{d^2}{d y^2 } R + \left(\sum_{i=1}^3  \frac{1}{y-y_i} \right) \odder{y} R +
 v(y) R =0~~,
\label{eqn:y}
\end{eqnarray}
where
\begin{eqnarray}
v(y) &=& \frac{1}{H(y)}\left[ \mu - \frac{y}{4} E - \sum_{i=1}^3
 \frac{\alpha_i^2 H'(y_i)}{y -y_i}\right]~,~~~~~
H(y) = \prod_{i=1}^3(y-y_i)~,\nonumber\\
\mu &=& \frac{E}{4} - \frac{3}{2} J(J+1) + \frac{3}{32} \left( \frac{2}{3}
 \frac{N_{\alpha} }{l} - Q_R\right)^2
\end{eqnarray}
and
\begin{eqnarray}
\alpha_1 &=& \pm \frac{1}{4} \left[ N_{\alpha} \left(p+q - \frac{1}{3l} \right) -Q_R
 \right]~~,\label{alpha 1}\\
\alpha_2 &=& \pm \frac{1}{4} \left[ N_{\alpha}  \left(p-q + \frac{1}{3l}\right) +Q_R
 \right]~~,\label{alpha 2}\\
\alpha_3 &=& \pm \frac{1}{4} \left[ N_{\alpha}
  \left(\frac{-2p^2+q^2+p\sqrt{4p^2-3q^2}}{q} - \frac{1}{3l}\right) -Q_R
 \right]~~.\label{alpha 3}
\end{eqnarray}
The exponents at the regular singularities are given by
$\pm \alpha_i$ at $y=y_i$ ($i=1,2,3$),
while
$-\lambda$ and $\lambda+2$ at $y=\infty$,
where we put 
\begin{eqnarray}
E=4 \lambda (\lambda+2).
\end{eqnarray}
It is convenient to transform the singularities from 
$\{y_1,y_2 ,y_3 ,\infty \}$ 
to $\{ 0,1,a =\frac{y_1 -y_3}{y_1 - y_2} , \infty \}$.
This is achieved by 
the transformation
\begin{eqnarray}
x= \frac{y -y_1}{y_2 - y_1} ~~
\end{eqnarray}
together with the rescaling
\begin{eqnarray}
R = x^{|\alpha_1|}   (1-x)^{|\alpha_2|}   (a-x)^{|\alpha_3|} h(x) ~, 
\end{eqnarray}
which transforms (\ref{eqn:y}) to the standard form of
Heun's equation:
\begin{eqnarray}
\frac{d^2}{dx^2}h(x) + \left(\frac{\gamma}{x} + \frac{\delta}{x-1} +
 \frac{\epsilon}{x-a}  \right) \frac{d}{dx} h(x) + \frac{\alpha \beta x
 - k}{x(x-1)(x-a)} h(x) = 0~~.
\end{eqnarray}
Here, Heun's parameters are given by
\begin{eqnarray}
&&\alpha = - \lambda + \sum_{i=1}^3 |\alpha_i| ~,~~~
\beta = 2 + \lambda + \sum_{i=1}^3 |\alpha_i|~,\nonumber\\
&&
\gamma = 1 + 2 |\alpha_1| ~,~~~
\delta = 1 + 2 |\alpha_2| ~,~~~
\epsilon =1 + 2 |\alpha_3| ~,
\end{eqnarray}
and $k$, which is called as the ``accessory'' parameter, is
\begin{eqnarray}
k &=& ( |\alpha_1| +  |\alpha_3| )( |\alpha_1| +  |\alpha_3| +1 ) - |\alpha_2
|^2 ~~\nonumber\\
&&+ a \left\{  ( |\alpha_1| +  |\alpha_2| )( |\alpha_1| +  |\alpha_2| +1 ) - |\alpha_3
|^2 \right\} - \tilde{\mu}
\end{eqnarray}
with
\begin{eqnarray}
\tilde{\mu} &=&
 - \frac{1}{y_1 - y_2} ( \mu - {y_1}\lambda(\lambda+2) ) ~~\nonumber\\
&=& \frac{p}{q} \left[ \frac{2}{3} (1-y_1)\lambda(\lambda+2) 
- J(J+1) + \frac{1}{16} \left(
 \frac{2}{3} \frac{N_{\alpha}}{l} -Q_R  \right)^2  \right]~,\\
a&=&\frac{1}{2}\left(
1+\frac{\sqrt{4p^2-3q^2}}{q}
\right)~.
\end{eqnarray}
Note that the parameter $a$ satisfies the inequality $a>1$
reflecting $p>q$.

The regularity of the eigenfunction in the range $0 \leq x \leq 1$, which
corresponds to $y_1 \leq y \leq y_2$ in the original variable, 
requires that $h(x)$ should be a Heun function in the sense of \cite{Ronveaux:1995};
$h(x)$ has the zero exponents at the both boundaries $x=0$ and $1$.
Simple such an example is a polynomial solution. 
The polynomial condition
on Heun's parameters
is non-trivial (see (27)),
although the necessary
condition is simply given by $\alpha \in {\mathbb Z}^-=\{ 0,-1,-2,-3, \cdots
\}$(or $\beta \in {\mathbb Z}^-$).
If we write the regular local solution around $x=0$ as
\begin{eqnarray}
h(x) &\simeq&  \sum_{n=0}^{\infty} c_n x^n ~,
\label{local solution}
\end{eqnarray}
then the radius of convergence is  normally 
$\min(1,a)$. 
Since $a>1$, the series will generically have
radius of convergence 1; 
it will therefore only represent a local
solution. 
If a certain condition on Heun's parameters
is satisfied,  the radius
of convergence is increased to $a$, 
so that  one can obtain a Heun function~\cite{Ronveaux:1995}. 
However, it is not easy to determine when the condition
is satisfied.
When the series breaks off at the $n$-th order,
(\ref{local solution}) becomes a polynomial solution of degree $n$.
Then, the coefficients $c_m$ satisfy a set of 
equations,
\begin{eqnarray}
-kc_0+a\gamma c_1&=&0~,\nonumber\\
P_mc_{m-1}-(Q_m+k)c_m+R_mc_{m+1}&=&0~~~~(m=1,2,\cdots,n-1)~,~~\nonumber\\
P_nc_{n-1}-(Q_n+k)c_n&=&0~,
\label{recursions}
\end{eqnarray}
where
\begin{eqnarray}
P_m&=&(m-1+\alpha)(m-1+\beta)~,\nonumber\\
Q_m&=&m\left(
(m-1+\gamma)(1+a) +a\delta+\epsilon
\right)~,\nonumber\\
R_m&=&(m+1)(m+\gamma)a~.
\end{eqnarray}

First,
we consider states with the quantum charges 
$\{ J, Q_R , N_{\alpha} \}$ corresponding to
the chiral operators of dual superconformal field theory.
Then, Heun's parameters are completely
fixed except for one parameter $\lambda$.
The basic chiral primary operators of mesons are given by the three fields
$\{\CS, \CL_+, \CL_- \}$
with charges given in Table 1  \cite{quiver,semiclassical string}.
\begin{table}[h]
 \begin{center}
  \begin{tabular}{|c|c|c|c|}
    \hline
Meson       &$J$    &$Q_R$    &$N_\alpha$    \\
    \hline
$\CS$       &$1$    &$2$    &$0$    \\
    \hline
$\CL_+$       &$\frac{p+q}{2}$    &$p+q-\frac{1}{3l}$    &$1$    \\
    \hline
$\CL_-$       &$\frac{p-q}{2}$    &$p-q+\frac{1}{3l}$    &$-1$    \\
    \hline
  \end{tabular}
 \end{center}
 \caption{Charge assignments for chiral primary operators.
 $N_\alpha$ corresponds to the U(1) flavor charge divided by $p$.}
\end{table}
One can read off these charges from the data of $Y^{p,q}$
quiver gauge theory.
Note that the $\CR$-charges of $\CL_\pm$
appear in the exponents $\alpha_{1,2}$
(see (\ref{alpha 1}) and (\ref{alpha 2}))
at the regular singularities $y_{1,2}$, respectively.
This is consistent with
the semi-classical analysis \cite{semiclassical string}.

When we choose the quantum charges 
$\{J,Q_R,N_\alpha\}$
as Table 1
together with $\lambda=\lambda_0$ in Table 2,
then Heun's parameters  $\alpha$ and $k$ vanish,
so that Heun's equation admits a constant solution
(a polynomial solution of degree 0) for each meson.
These solutions represent ground states with fixed 
$\{J,Q_R,N_\alpha\}$\fnote{$\flat$}{
These constant solutions correspond to the holomorphic functions
on the 
Calabi-Yau cone $C(Y^{p,q})$ examined in \cite{BHOP}.
In fact,
the $y$-dependence of the holomorphic functions,
$(y-y_1)^{|\alpha_1|}(y-y_2)^{|\alpha_2|}(y-y_3)^{|\alpha_3|}$,
can be read off from the equation (20).
We thank Christopher Herzog for
pointing out \cite{BHOP}.
}.
\begin{table}[h]
 \begin{center}
  \begin{tabular}{|c|c|}
    \hline
 Meson      &$\lambda_0$    \\
    \hline
 $\CS$      &$\frac{3}{2}$    \\
    \hline
 $\CL_+$      &$\frac{p}{4}(
 \frac{2p-\sqrt{4p^2-3q^2}}{q}+3
 )$    \\
    \hline
 $\CL_-$      &$\frac{p}{4}(
 \frac{-2p+\sqrt{4p^2-3q^2}}{q}+3
 )$     \\
    \hline
  \end{tabular}
 \end{center}
  \caption{The corresponding ground energy $E_0=4\lambda_0(\lambda_0+2)$.}
\end{table}

The conformal dimension $\Delta$
of the dual operator
 is related to the
ground energy $E_0$ by the formula \cite{AdS/CFT}\cite{Gubser:1998vd},
\begin{eqnarray}
\Delta=-2+\sqrt{4+E_0}~.
\label{conformal dimension}
\end{eqnarray}
Using Table 2,
we obtain the equality $\Delta=\frac{3}{2}Q_R$.
In this way,
we have reproduced the BPS condition
of the dual operator from the gravity side (see also \cite{BHOP}).
This is consistent with
the AdS/CFT
correspondence.

Next, let us consider
the first excited state with $\lambda=\lambda_0+1$ and
$\{J,Q_R,N_\alpha\}$ in Table 1.
We find that it exists and
the corresponding Heun function $h_1(x)$
is given by the polynomial solution of degree 1;
\begin{eqnarray}
h_1(x)=\left\{
  \begin{array}{ll}\displaystyle
1-\frac{2p+3q+\sqrt{4p^2-3q^2}}{q+\sqrt{4p^2-3q^2}} x      
&~~~\mbox{for}~~\CS ~,    \\
\displaystyle
1+\frac{2p-3q-pq+(1+p)\sqrt{4p^2-3q^2}}{(1+p)(-q+\sqrt{4p^2-3q^2})}
(x-1)
&~~~\mbox{for}~~\CL_+ ~,    \\
\displaystyle
1-\frac{2p+3q+pq+(1+p)\sqrt{4p^2-3q^2}}{(1+p)(q+\sqrt{4p^2-3q^2})}
x       
&~~~\mbox{for}~~\CL_-  ~.   \\
  \end{array}
\right.
\label{h_1}
\end{eqnarray}
The existence of the higher excited states with 
$\lambda=\lambda_0+n$ ($n=2,3,\cdots$) is indicated
by numerical simulation.
For them, Heun functions are not polynomials,
although the parameter $\alpha$ is a negative integer $-n$.
The difference $\Delta_n-\frac{3}{2}Q_R$
is equal to $2n$,
where
$\Delta_n=-2+\sqrt{4+E_n}$ and 
$E_n=4(\lambda_0+n)(\lambda_0+n+2)$.

We find more general solutions
with 
Heun's parameters summarized in Table 3.
\begin{table}[h]
 \begin{center}
  \begin{tabular}{|c||c|c|c|c|c|}
    \hline
       &$\alpha$    &$\beta$    &$\gamma$    &$\delta$    &$\epsilon$    \\
    \hline\hline
$A$       &$-n$    &$2+3N+n$    &$1+N$    &$1+N$    &$1+N$  \\
    \hline
$B$       &$-n$    &$2+n+Np+\frac{Np(2p+q-\sqrt{4p^2-3q^2})}{2q}$    &$1$    &$1+Np$    
&$1+\frac{Np(2p+q-\sqrt{4p^2-3q^2})}{2q}$   \\
    \hline
$C$       &$-n$    &$2+n+Np+\frac{Np(-2p+q+\sqrt{4p^2-3q^2})}{2q}$    &$1+Np$    &$1$    
&$1+\frac{Np(-2p+q+\sqrt{4p^2-3q^2})}{2q}$ 
   \\
    \hline
  \end{tabular}
\smallskip
  
\begin{tabular}{|c||c||c|}
    \hline
       &$k$    &$\lambda$    \\
    \hline\hline
$A$           
&$-\frac{n(2+n+3N)(2p+3q+\sqrt{4p^2-3q^2})}{6q}$    &$n+\frac{3}{2}N$    \\
    \hline
$B$           
&$-n\frac{6Np^2+2p(2+n+3Nq)+(2+n)(3q+\sqrt{4p^2-3q^2})}{6q}$    
&$n+\frac{p}{4}(\frac{2p-\sqrt{4p^2-3q^2}}{q}+3)N$    \\
    \hline
$C$        
&$-n\frac{3Npq+3q(n+2)+4p+8n+(2+n+3Npq)\sqrt{4p^2-3q^2}}{6q}$    
&$n+\frac{p}{4}(\frac{-2p+\sqrt{4p^2-3q^2}}{q}+3)N$    \\
    \hline
  \end{tabular}
 \end{center}
 \caption{Heun's parameters and $\lambda$ for three-types of states,
 $A$, $B$ and $C$
 ($N=1,2,\cdots$ and $n=0,1,2,\cdots$).
 }
\end{table}
These solutions
reduce to those given above
if we set $N=1$.
Even if $N>1$,
one can obtain the similar 
results to those found in the case of $N=1$.
Especially, when we set $n=0$,
the states $A$, $B$ and $C$
represent those dual to $N$ mesonic chiral operators
corresponding to $\CS$, $\CL_+$ and $\CL_-$, respectively.
The quantum charges for $N$ meson operators are
obtained by
multiplying charges in Table 1 and 2 by $N$,
\begin{eqnarray}
J\to NJ~,~~~
Q_R\to NQ_R~,~~~
N_\alpha\to NN_\alpha~~~
\mbox{and} ~~~
\lambda_0\to N\lambda_0~.
\end{eqnarray}

Finally,
taking $N$ large,
we find that
the  Heun functions take the form
$h_n(x)\simeq n$-polynomial $+~{\mathcal O}(1/N)$,
and in the limit $N\to \infty$,
they tend to the following polynomials;
\begin{eqnarray}
h_n(x)\to\left\{
  \begin{array}{ll}\displaystyle
~~h_1(x)^n    
  &~~~\mbox{for}~~A     \\
~~
  x^n     &~~~\mbox{for}~~B     \\
~~
  (1-x)^n     &~~~\mbox{for}~~C     \\
  \end{array}
\right.
\end{eqnarray}
where $h_1(x)$ is defined in (\ref{h_1}).
These polynomials are derived by
using $1/N$-expansion of three-term recursions (\ref{recursions})\fnote{$\diamond $}{
For $B$, we use three-term recursions around $x=1$
instead of $x=0$
for the consistent limit $N\to\infty$.
}.
The near BPS states with $n\ll N$
correspond
to the near BPS geodesics in
\cite{semiclassical string}.

\medskip

In this letter, we studied
the spectrum of the scalar Laplacian on $Y^{p,q}$ corresponding to
mesonic chiral operators of the dual superconformal field theories.
The states dual to the baryonic chiral operators
will be extracted in the similar manner.
In \cite{Ypq}, inhomogeneous toric Sasaki-Einstein manifolds in
arbitrary odd-dimensions have been constructed.
It is also interesting to examine the spectrum of the scalar Laplacian
on seven-dimensional toric Sasaki-Einstein manifolds
and the 3-dimensional dual gauge theories.
For a large class of Sasaki-Einstein spaces,
their volumes are calculated in \cite{BH}.
It is interesting to derive the volume of the non-toric 
3-Sasakian
manifolds constructed 
in \cite{Sakaguchi:2004xp}.
We hope to report this point in the future.
After \cite{Hashimoto:2004ks},
many compact Sasaki-Einstein manifolds \cite{Lpqr}
have been derived from
AdS-Kerr black holes.
It is important to
clarify the spectrum on these manifolds.
These are left for future investigations.

\vspace{4mm}

We are grateful to Yoshitake Hashimoto
and
Sean Hartnoll
for discussions
and useful comments.
This work is supported by the 21 COE program
``Constitution of wide-angle mathematical basis focused on knots"
and in part by the Grant-in
Aid for scientific Research (No.~17540262 and No.~17540091)
from Japan Ministry of Education.


\end{document}